\documentclass[]{emulateapj}

\usepackage{graphicx,times}

\begin{document}

\title{Frequency identification and asteroseismic analysis of the red giant KIC 9145955: fundamental parameters and helium core size}
\author{Xinyi Zhang\altaffilmark{1,2,3},Tao Wu\altaffilmark{1,2,4},Yan Li\altaffilmark{1,2,3,4}}
\email{zhangxinyi@ynao.ac.cn; wutao@ynao.ac.cn; ly@ynao.ac.cn}

\altaffiltext{1}{Yunnan Observatories, Chinese Academy of Sciences, 396 Yangfangwang, Guandu District, Kunming, 650216, P. R. China; zhangxinyi@ynao.ac.cn; wutao@ynao.ac.cn; ly@ynao.ac.cn}
\altaffiltext{2}{Key Laboratory for Structure and Evolution of Celestial Objects, Chinese Academy of Sciences, 396 Yangfangwang, Guandu District, Kunming, 650216, P. R. China}
\altaffiltext{3}{University of Chinese Academy of Sciences, Beijing 100049, China}
\altaffiltext{4}{Center for Astronomical Mega-Science,Chinese Academy of Sciences,20A Datun Road,Chaoyang District,Beijing 100012,China}

\begin{abstract}
  We have analyzed 18 quarters of long-cadence data of KIC 9145955 provided by \emph{Kepler}, and extracted 61 oscillation frequencies from these high precision photometric data. The oscillation frequencies include 7 $l = 0$ modes, 44 $l = 1$ modes, 7 $l = 2$ modes, and 3 $l = 3$ modes. We identify $l = 0$ modes as p modes and $l = 2$ modes as p-dominated modes. For $l = 1$ modes, all of them are identified as mixed modes.
  These mixed modes can be used to determine the size of the helium core. We conduct a series of asteroseismic models and the size of the helium core is determined to be $M_{\rm He}$ = 0.210 $\pm$ 0.002  $M_{\odot}$ and $R_{\rm He}$ = 0.0307 $\pm$ 0.0002 $R_{\odot}$.
  Furthermore, we find that only the acoustic radius $\tau_{0}$ can be precisely determined with the asteroseismic method independently.
  The value of $\tau_{0}$ is determined to be 0.494 $\pm$ 0.001 days. By combining asteroseismic results and spectroscopic observations, we obtain the best-fitting model. The physical parameters of this model are
  $M$ = 1.24 $M_{\odot}$, $Z$ = 0.009, $\alpha$ = 2.0, $T_{\rm eff}$ = 5069 K, $\log g$ = 3.029, $R$ = 5.636 $R_{\odot}$, and $L$ = 18.759 $L_{\odot}$.
  In addition, we think that the observed frequency F39 (96.397 $\mu$Hz) is more appropriate to be identified as a mixed mode of the most p-dominated.
\end{abstract}

\keywords{asteroseismology-stars:individual(KIC 9145955)-stars:oscillations-stars:solar-type}

\section{Introduction}
Space missions such as CoRoT (Baglin et al. 2006) and \emph{Kepler} (Borucki et al. 2008) have observed a lot of high precision
oscillation frequencies of pulsating stars. Asteroseismology has made many achievements on probing the internal structure of stars. When hydrogen is
exhausted in the center of a solar-like star, it will leave the main sequence and become a red giant. A hydrogen burning shell above the helium core
supports the luminosity of the star at this stage. The observed pulsation spectra of red giants exhibit the signature of solar-like oscillations
(De Ridder et al. 2009; Bedding et al. 2010; Huber et al. 2010). Solar-like oscillations are excited stochastically by turbulence in the stellar convective envelope. The restoring force of solar-like oscillations is the pressure gradient. These oscillations have the nature of acoustic waves and are known as p modes. The oscillation spectra of evolved stars also show gravity modes (g modes) whose dominant restoring force is buoyancy. In sub-giants and red giants, g modes may have higher frequencies than stars on the main sequence as a result of their compact cores and are in the frequency range of p modes. This produces mixed modes, which behaves as p-mode signature in the stellar envelope and g-mode signature in the stellar core. The mixed modes in red giants were first identified by Bedding et al. (2010), which were in agreement with the theoretically prediction prior to observations (e.g., Dziembowski et al. 2001; Christensen-Dalsgaaed 2004; Dupret et al. 2009). With the detection of mixed modes in the fields of \emph{CoRoT} and \emph{Kepler} (e.g., Beck et al. 2011; Mosser et al. 2011), asteroseismology has been successfully applied to probe the core regions of red giants. Beck et al. (2011) first measured the period spacings of mixed modes in a red giant observed by \emph{Kepler} space telescope. Subsequently, Bedding et al. (2011) and Mosser et al. (2011) confirmed that hydrogen-shell burning giants and core-helium burning stars can be distinguished by different period spacings of mixed modes. Stello et al. (2013) measured the period spacings of mixed modes of 13000 \emph{Kepler} targets to classify these stars into various populations, such as the red giant branch, the helium-core burning red clump and the secondary clump. Montalban et al. (2013) proposed a test of overshooting during the main-sequence and core He-burning phase by using period spacings of mixed modes in red giants. Gai \& Tang (2015) suggested that the asymptotic g-mode period spacing of red giants is able to distinguish stars in different positions of the RGB bump which are in different evolutionary stages and structure of cores but have similar fundamental parameters. In addition, individual oscillation frequencies of a particular red giant are used to constrain stellar fundamental parameters, such as done by Kallinger et al. (2008), Di Mauro et al. (2011), Deheuvels et al. (2012), Lillo-Box et al. (2014), Quinn et al. (2015), Di Mauro et al. (2016) and Li et al. (2017).

KIC 9145955 is a good example of a bright ($m_{\rm V}$ = 10.05) low-luminosity RGB star with very high signal-noise-ratio of the individual frequencies and does not show complicated power density spectrum caused by rotational splitting. Therefore, KIC 9145955 was mentioned in several papers.
KIC 9145955 was first identified as a red giant by Bedding et al. (2011), and continuously observed by \emph{Kepler} for four years. Hekker et al. (2014) presented the fourier power density spectrum of KIC 9145955. Datta et al. (2015) determined the period spacing of KIC 9145955 to be 76.98 $\pm$ 0.03 s by an automated procedure. In addition, Vrard et al. (2016) measured the period spacings of g modes for more than 6100 red giants in the \emph{Kepler} fields and gave the period spacing of KIC 9145955 to be 77.1 s, as well as the large frequency separation to be 11.03 $\mu$Hz. The oscillation frequencies of KIC 9145955 were firstly exhibited by using Bayesian peak bagging analysis (Corsaro et al. 2015a). Takeda et al. (2016) selected 48 \emph{Kepler} red giants, including KIC 9145955, which were selected from the list of Mosser et al. (2012), and performed spectroscopic observations in July 2015 by using Subaru/High Dispersion Spectrograph (HDS). Based on individual pulsational frequencies and spectroscopic data, P{\'e}rez Hern{\'a}ndez et al. (2016) obtained asteroseismic parameters of 19 Kepler red giants, including KIC 9145955, such as masses and radii. However, P{\'e}rez Hern{\'a}ndez et al. (2016) did not model all $l = 1$ frequencies but period spacing of $l = 1$ frequencies instead. In addition, Hekker \& Christensen-Dalsgaard (2017) made a thorough review on asteroseismology of red giants, using KIC 9145955 as an example and presenting its power density spectrum in three figures.

In this paper, we use individual frequencies include $l = 1$ modes to conduct asteroseismic analysis for KIC 9145955 to restrict its global parameters and internal structure. In section 2, we present oscillation frequency extraction and spectroscopic data. In section 3, we describe our asteroseismic model in details. Input physics, model grids and model fittings are described in section 3.1, section 3.2, and section 3.3, respectively. In section 4, we present and discuss our results of model calculations. The frequency identification is described in section 4.1. We analyze how to select the best-fitting model in section 4.2, and reduce the offset between observed and calculated p modes in section 4.3. We compare our results with previous work in section 5. Finally, we summarize our work in section 6.

\section{Frequency analysis and observations}

\subsection{Oscillation frequency}
\emph{Kepler} Asteroseismic Science Operations Center(KASOC) data base (http://kasoc.phys.au.dk/) provides us with the observational data. \emph{Kepler} data are organized into quarters. Long-Cadence quarters are marked as Qn and Short-Cadence quarters are marked as Qn.m (Murphy 2012). The photometric data of KIC 9145955 are available in 18 quarters (Q0-Q17) of Long-Cadence modes (29.42 minutes) and 1 quarter (Q4.3) of Short-Cadence modes (58.85 seconds).
We only use the Long-Cadence time series data of all 18 quarters which were observed from May 2009 to May 2013.

We use the software package Period04 (Lenz \& Breger 2005) to extract individual frequencies from the photometric data. Our method of extracting frequencies is based on the asymptotic theory of stellar oscillations. The duration of the observations reached 1470.46 days, which leads to an excellent frequency resolution of 0.079 $\mu$Hz. In the power density spectrum, the frequency resolution is smaller than the separation between adjacent potential frequencies. Most of modes can be easily picked out from the power density spectrum. For example, all $l = 0$ modes are clearly distinguish from other modes. Therefore, they are all well-resolved. Then, we use Period04 to extract individual frequencies. But we find that there could be more than one peak for a potential frequency, such as $l = 0$ modes. For this case, we extract the highest peak by using Period04 to obtain the frequency.
The frequencies we extracted from the power density spectrum are consistent with the asymptotic theory. Altogether 61 frequencies with the signal-to-noise ratio (S/N) $>$ 4 are listed in Table 1. The top panel of Figure 1 shows the power density spectrum of KIC 9145955 and the oscillation frequencies are marked by colored lines.

The observed frequencies of KIC 9145955 have been published by Corsaro et al. (2015a), who obtained 70 frequencies. In our work, 10 frequencies, including 2 $l = 0$ modes, 5 $l = 1$ modes, 2 $l = 2$ modes and 1 $l = 3$ modes, are excluded due to their low signal-to-noise ratios. Moreover, we extract 1 more $l = 1$ mode F42 = 98.840 $\mu$Hz that was not mentioned in the work of Corsaro et al. (2015a). In our work, there are 8 frequencies, including 3 $l = 0$ frequencies, 1 $l = 1$ frequency, and 4 $l = 2$ frequencies, in disagreement with the results of Corsaro et al. (2015a) within error bars which are derived from Period04. However, among the 8 frequencies, the maximum deviation between our results and the results of Corsaro et al. (2015a) is no more than 0.074 $\mu$Hz. Therefore our frequencies are reliable, especially for the frequencies of $l = 1$ modes which are almost identical with the results of Corsaro et al. (2015a).

 In the top panel of Figure 1, the power density spectrum of KIC 9145955 does not show the rotational splitting of frequency. Corsaro et al. (2015a) also presented that for KIC 9145955, there is no evidence for rotational splitting. This fact makes peakbagging easier. But we have to admit that there is a selection effect: maybe this star is viewed nearly pole-on. We present the oscillation frequencies extracted from the power density spectrum (top panel of Figure 1) in the form of a frequency \'{e}chelle diagram in Figure 2 and a period \'{e}chelle diagram in Figure 3. The frequencies with the spherically harmonic degree $l$ = 0, 2, and 3 show vertical ridges in Figure 2. In Figure 2 and Figure 3, it can be seen that only $l = 1$ modes show properties of the mixed modes. In Figure 3, we first identify the mode farthest from the vertical dashed line of each line as the most p-dominated. We should note that both p- and g-dominated modes are the mixed modes with both p-mode and g-mode nature. The mode identification will be discussed in details in section 4.1. In the bottom panel of Figure 1, each filled cycle represents a period spacing between two adjacent $l = 1$ modes which correspond to adjacent blue lines in the top panel of Figure 1. It should be noticed in the bottom panel of Figure 1 that two g-dominated mixed modes were not observed. From the frequency \'{e}chelle diagram in Figure 2 and the period \'{e}chelle diagram in Figure 3, we obtain the values of the large frequency separation and period spacing as $\Delta \nu$ = 11.065 $\mu$Hz and $\Delta P$ = 77.01 s, respectively.

\subsection{Spectroscopic observations}
Takeda et al. (2016) conducted spectroscopic observations for 48 giants in the \emph{Kepler} fields on July 3, 2015. The red giant KIC 9145955 is one of those targets. Their results are: $T_{\rm eff}$ = 4943 K, $\log g$ = 2.85, [Fe/H] = -0.34. Based on the APOKASC Catalogue (Pinsonneault et al. 2014), P{\'e}rez Hern{\'a}ndez et al. (2016) obtained effective temperature being $T_{\rm eff}$ = 4925 $\pm$ 91 K, surface gravity being
$\log g$ = 3.04 $\pm$ 0.11 dex, and metallicity being [Fe/H] = -0.32 $\pm$ 0.03 dex, respectively. These parameters are listed in Table 2.

\section{Asteroseismic model}

\subsection{Input physics}
Our stellar models are calculated by the Modules for Experiments in Stellar Astrophysics (MESA; Paxton et al. 2011, 2013) of version 6596 with the following input physics: the OPAL equation of state tables (Rogers \& Nayfonov 2002) and the OPAL opacity tables (Iglesias \& Rogers 1996) supplemented by Ferguson et al. (2005). We adopt the Eddington grey $T - \tau$ integration as the atmospheric boundary condition. The mixing-length theory (MLT) of B\"{o}hm-Vitense (1958) is used to deal with convection. The MESA module called pulse, which is based on the Aarhus adiabatic pulsation code ADIPLS (Christensen-Dalsgaard 2008), supports calculations of stellar evolution and oscillation frequencies in our case. In addition, we do not consider rotation, diffusion, convective overshooting, and magnetic field in our calculations.

\subsection{Model grids}
A grid of theoretical stellar models are computed, the initial mass $M$ ranging from 1.20 $M_{\odot}$ to 1.30 $M_{\odot}$ with a step of 0.01 $M_{\odot}$, the initial metallicity $Z$ ranging from 0.006 to 0.015 with a step of 0.001, and the mixing-length parameter $\alpha$ ranging from 1.8 to 2.2 with a step of 0.1. The helium abundance function $Y = 0.245 + 1.54Z$ (e.g., Dotter et al. 2008; Thompson et al. 2014) is adopted as the initial input value. Each star is computed from the pre-main-sequence to the red giant stage. Since the luminosity changes greatly while the effective temperature changes slightly when a star evolves along the red giant branch, we use the luminosity as the range of frequency calculation box: 15$L_{\odot}$ $<$ L $<$ 23$L_{\odot}$. When a star evolves into the frequency calculation box on the Hertzsprung-Russel diagram, a smaller time step is adopted to slow the evolution, and frequencies of oscillation modes with $l$ = 0, 1, and 2 are calculated for every theoretical model falling into the frequency calculation box.

\subsection{Model fittings}
We try to find the best-fitting model that reflects the realistic structure of KIC 9145955 through the goodness-of-fit functions, which are defined as:
\begin{equation}
\chi^{2}_{\rm all}=\frac{1}{K}\sum(\nu_{k,l}^{\rm obs}-\nu_{k,l}^{\rm mod})^{2}
\end{equation}
and
\begin{equation}
\chi^{2}_{l=1}=\frac{1}{K}\sum(\nu_{k,l=1}^{\rm obs}-\nu_{k,l=1}^{\rm mod})^{2}.
\end{equation}
Here, all observed frequencies are used in Equation (1), while only the $l = 1$ observed frequencies are used in Equation (2). In above equations, $\nu_{k,l}^{\rm obs}$, $\nu_{k,l=1}^{\rm obs}$, $\nu_{k,l}^{\rm mod}$, and $\nu_{k,l=1}^{\rm mod}$ correspond to the observed frequencies and our model eigenfrequencies, respectively. K is the total number of the observed oscillation frequencies. We do not include the $l = 3$ observed frequencies in our calculations for the values of $\chi^{2}$, because this will greatly increase complexity in the frequency calculations. Although we list the errors derived from Period04, we do not think that it is necessary to include these errors in the goodness-of-fit functions for the following reasons: (1) The frequency errors obtained from Period04 are not accurate. Fu et al. (2013) found that errors derived from Period04 were underestimated by more than 70 percent compared with the errors derived from the Monte Carlo simulations. (2) Rotation is definitely present in the star and contributes to the nonradial modes split into 2$l$ + 1 different frequencies. We have not seen the rotational splitting due to a selection effect which we have mentioned in section 2.1. For $l = 1$ frequencies, three frequencies will form a triplet because of rotational splitting, and we can not distinguish whether the extracted frequency is at the center or the edge of the triplet. This makes frequency determinations more or less uncertain.

\section{Results and Discussions}

\subsection{Frequency identification}
In order to calculate the value of $\chi^{2}$, we match the observed frequencies with their corresponding calculated frequencies. For $l = 0$ modes, the observed frequencies are easily matched with the calculated ones. For $l = 2$ modes, the observed frequencies are p-dominated mixed modes. The inertias of $ l = 2$ modes are very high and only the frequencies which are close to the pure acoustic $l = 2$ modes can be detected (Hekker \& Christensen-Dalsgaard 2017). Therefore, we first find the calculated frequency with the minimal mode inertia $I_{kl}$ and then match it with the observed frequency. The mode inertia is defined as (Christensen-Dalsgaard 2003)
\begin{equation}
I_{kl}=\frac{4\pi\int_{0}^{R}[\xi_{r}(r)^{2}+l(l+1)\xi_{h}(r)^{2}]\rho_{0}r^{2}dr}{M[\xi_{r}(R)^{2}+l(l+1)\xi_{h}(R)^{2}]},
\end{equation}
where M is the total mass of the star, $\rho_{0}$ is the local density, the quantities $\xi_{r}(r)$ and $\xi_{h}(r)$ are radial and horizontal displacements, respectively.
For $l = 1$ modes, however, all of the observed frequencies are mixed modes. We compute a p-mode energy ratio $E_{p}/E_{kl}$ for each $l = 1$ mode. We define a mixed mode with the highest value of $E_{p}/E_{kl}$ in a large frequency separation as the most p-dominated mode. The $E_{p}/E_{kl}$ is the ratio between the p-mode oscillation energy and the total oscillation energy of the same frequency.
The oscillation energy $E_{kl}$ is calculated by the following equation (Christensen-Dalsgaard 2003)
\begin{equation}
E_{kl}=4\pi\int_{0}^{R}[\xi_{r}(r)^{2}+l(l+1)\xi_{h}(r)^{2}]\rho_{0}r^{2}dr.
\end{equation}
We obtain the total oscillation
energy $E_{kl}$ by integrating Equation (4) from the center to the stellar surface. To obtain the p-mode oscillation energy $E_{p}$, we need to determine
the p-mode propagation region. According to the linear oscillation theory, the propagation region of a p mode satisfies the condition: $|\omega|$ $>$ $|N|$ and $|\omega|$ $>$ $S_{l}$, while that of a g mode satisfies the condition: $|\omega|$ $<$ $|N|$ and $|\omega|$ $<$ $S_{l}$. The symbol $S_{l}$ is the characteristic acoustic frequency (Christensen-Dalsgaard 2003)
\begin{equation}
S_{l}^{2}=\frac{l(l+1)c_{s}^{2}}{r^{2}}
\end{equation}
and $N$ is the buoyancy frequency (Christensen-Dalsgaard 2003)
\begin{equation}
N^{2}=g(\frac{1}{\Gamma_{1}p}\frac{dp}{dr}-\frac{1}{\rho}\frac{d\rho}{dr}),
\end{equation}
where $p$ is the local pressure, $c_{s}$ is the adiabatic sound speed, and $\Gamma_{1}$ is the adiabatic exponent.

First, we conduct model calculations and fittings by following the frequency identification method which we have mentioned in section 2.1.
When the mixing-length is fixed as 2.0, we list several candidate models in Table 3. These models are named as Model A to Model H. We present the \'{e}chelle diagrams for Model E and Model F of Table 3 in Figure 4. Table 3 shows that Model F has a lower value of $\chi^{2}_{l=1}$ than Model E. However, Figure 4 shows that the matching of Model E between $l = 1$ observed and computed frequencies is better than Model F. We also notice that an observed mode of the most p-dominated might be matched with a calculated g-dominated mixed mode, and an observed g-dominated mixed mode might be matched with a calculated mode of the most p-dominated. In Figure 5, we display the p-mode oscillation energy ratio of Model E. Figure 5 shows that the calculated frequency 96.300 $\mu$Hz, which corresponds to the observed frequency F39 = 96.397 $\mu$Hz, has a higher value of $E_{p}/E_{kl}$. But the observed frequency F39 = 96.397 $\mu$Hz has been identified as a g-dominated mixed mode.

Then, we choose an alternative frequency matching scheme, i.e., to match F37 = 96.931 $\mu$Hz with a calculated frequency of relatively
lower value of $E_{p}/E_{kl}$ and to match F39 = 96.397 $\mu$Hz with a calculated frequency of higher p-mode oscillation energy ratio.
Our results of model calculations are listed in Table 4. We display in Figure 6 the \'{e}chelle diagrams for Model D of Table 3 in panel (a) and for Model U of Table 4 in panel (b). These two models have similar minimal values of $\chi^{2}_{l=1}$. However, there are great differences when comparing panel (a) and panel (b) in Figure 6. The model in panel (a) has a minimal value of $\chi^{2}_{l=1}$, but the differences between the observed and calculated $l = 1$ frequencies are much larger than the model exhibited in panel (b). These facts clearly justify the matching scheme used for models in Table 4. Therefore, we think it is more appropriate to identify F39 = 96.397 $\mu$Hz as the most p-dominated mode.

\subsection{The best-fitting model}
In the frequency calculation box, every evolutionary track has a minimal value of $\chi^{2}$. We obtain all together 550 minima of $\chi^{2}$ in the
frequency calculation box and pick out the lowest value from these minima. The lowest values of $\chi^{2}_{\rm all}$ and $\chi^{2}_{l=1}$ are 0.0433 and 0.0381, respectively. Several models with the minimal values of $\chi^{2}_{\rm all}$ and $\chi^{2}_{l=1}$ near the above lowest values are also listed
in Table 4.

Table 4 shows that the values of $\chi^{2}_{\rm all}$ are larger than the values of $\chi^{2}_{l=1}$ for every model. The reason is that the surface
effect causes the frequencies offset, especially for the frequencies of $l = 0$ and $l = 2$ modes. In addition, Model U has a minimal $\chi^{2}_{l=1}$
value and Model Q has a minimal $\chi^{2}_{\rm all}$ value among all 550 calculated models. How can we select the best-fitting model from them ?

In Figure 7, we exhibit the $\chi^{2}_{\rm all}$ as a function of different model parameters. Panel (a) shows that all 550 evolutionary tracks converge to
a minimal value when the model parameter is the acoustic radius $\tau_{0}$. On the other hand, with other parameters, like surface gravity $\log g$,
stellar radius $R$, and stellar luminosity $L$, the evolutionary tracks do not converge but occupy a wider range as shown in panel (b)-(d).
Wu et al. (2016) also discovered that the global parameters that could be best measured by $\chi^{2}_{\rm \nu}$-matching method was the acoustic radius for a main-sequence star KIC 6225718. The acoustic radius is the sound travel time from the core to the surface of a star, defined by (Aerts et al. 2010)
\begin{equation}
\tau_{0} = \int_{0}^{R}\frac{dr}{c_{s}(r)}.
\end{equation}
Here $c_{s}$ represents the adiabatic sound speed and $R$ is the stellar radius. Usually $c_{s}$ is much larger in the helium core but becomes quite smaller in the stellar envelope. As a result, the acoustic radius $\tau_{0}$ is more suitable to describe the properties of the stellar envelope.
It can be noticed in Panel (a) of Figure 7 that all evolutionary tracks converge to a minimal value of $\chi_{\rm all}^{2}$ when the acoustic radius
is about 0.495 days.

Panel (b) of Figure 6 shows that the calculated frequencies of $l = 1$ modes are highly consistent with observations. Therefore, we can use the value
of $\chi^{2}_{l=1}$ to pick out the best-fitting model. We exhibit propagation diagram of Model N listed in Table 4 with $\chi^{2}_{l=1}$ = 0.0385 in
Figure 8 to show clearly properties of the calculated mixed modes. Figure 8 shows the profile of the buoyancy frequency $N$ and characteristic
frequency $S_{1,2}$, the vertical dashed line line marked by $X_{H} = 0.01$ indicating the helium core boundary where the hydrogen mass fraction
is about 0.01. There is a peak for the buoyancy frequency near the helium core boundary, which is caused by the change of density. In the red giant
phase, a hydrogen burning shell operates above the helium core, resulting in a sharp increase of hydrogen abundance and thus a corresponding decrease
of density. We plot a density line with a logarithmic ordinate in the graph of hydrogen abundance of Model N in Figure 9. It can be seen clearly that
the slope of density changes at the boundary of the helium core. Figure 10 shows the scaled displacement eigenfunctions of six $l = 1$ eigenfrequencies of Model N. The information of these frequencies are listed in Table 5. The ordinate $r\rho^{\frac{1}{2}}\xi_{r}$ is related to the oscillation energy. All
of $l = 1$ frequencies are seen to be the mixed modes. They have p-mode features in the stellar envelope and g-mode features in the stellar core.
We can also find that the displacement becomes smaller at the helium core boundary and increases when the wave goes across the boundary. This phenomenon
can be explained by the fact that when the internal gravity wave propagates to the hydrogen burning shell near the helium core boundary, part of it is
reflected and the rest is refracted. The mixed modes can propagate into the stellar core and carry information about the stellar helium core. Therefore,
we can precisely determine the helium core parameters of KIC 9145955 by using the observed mixed modes. We notice that all candidate models listed in
Table 4 have similar values of $\chi^{2}_{l=1}$, and the mass and radius of the helium core are highly consistent with all of the considered models.
These facts indicate that those models listed in Table 4 reflect a common feature of helium core for the considered star KIC 9145955.

Because the best-fitting model can not be selected based on the observed oscillation frequencies of the considered star, we have to use more parameters
that can be observed by other means. We notice from Table 4 that the initial metallicity $Z$ increases as the stellar mass $M$ becomes bigger for the
candidate models with fixed values of the mixing-length parameter $\alpha$. It is certain in discussions of Section 2.2 that the spectroscopic
observations can help us to constrain the stellar metallicity. The spectroscopic values of [Fe/H] listed in Table 2 are -0.34 (Takeda et al. 2016),
and -0.32 $\pm$ 0.03 (P{\'e}rez Hern{\'a}ndez et al. 2016). Obviously, this star is metal-poor and its heavy element abundance should be lower than the
solar value.
The metallicity [Fe/H] is defined as:
\begin{equation}
\rm [Fe/H]=\lg(\frac{Z}{X})_{star}-\lg(\frac{Z}{X})_{\odot},
\end{equation}
where we adopt $(Z/X)_{\odot}$ = 0.0245 (Grevesse and Noels 1993) and the value of [Fe/H] from P{\'e}rez Hern{\'a}ndez et al. (2016). As a result, the value of $(Z/X)_{\rm star}$ for KIC 9145955 should be in the range of 0.0109 $\sim$ 0.0126.

The majority of the observed frequencies are mixed modes, and they are matched well with the computed mixed modes. Therefore, we use the atmospheric metallicity [Fe/H] and the value of $\chi^{2}_{l=1}$ to find the best-fitting model. To select the best-fitting model, two criteria are taken into consideration by us: one is the model with relatively low value of $\chi^{2}_{l=1}$; the other is the value of $(Z/X)_{\rm star}$ being
consistent with the spectroscopic observations. As shown in Table 4, Model U has the minimal value of $\chi^{2}_{l=1}$ (0.381), but its value of
$(Z/X)_{\rm star}$ (0.0213) is too high compared with the observationally acceptable values. For Model N, however, its value of $\chi^{2}_{l=1}$ is 0.0385, a little bit higher than Model U but considerably lower than other models. On the other hand, its value of $(Z/X)_{\rm star}$ is 0.0139,
only a little bit higher than the observational range. Therefore, we consider Model N as the best-fitting model to KIC 9145955.

\subsection{Surface effect}
In Figure 6, there are systematic offsets between measured and computed frequencies of $l = 0$ and $l = 2$ modes. These frequency shifts are independent of the spherically harmonic angular degree ($l$) of the modes and significant for higher order p-mode frequencies. These effects are known as the surface
effect, caused by imperfect modelling of the uppermost stellar layers (Christensen-Dalsgaard et al. 1988, 1996; Dziembowski et al. 1988;
Christensen-Dalsgaard \& Thompson 1997). Several empirical corrections for the near-surface offset have been proposed to reduce these frequency discrepancies, e.g., Kjeldsen et al. (2008), Ball \& Gizon (2014), Sonoi et al. (2015), Ball et al. (2016), Ball \& Gizon (2017). Kjeldsen et al. (2008) used data for the sun to derive an empirical correction. This method has subsequently been applied widely to other stars (e.g., Christensen-Dalsgaard et al. 2010, Metcalfe et al. 2010, Deheuvels et al. 2012). We adopt the surface corrections described by Brand{\~a}o et al. (2011). For radial modes and
nonradial modes, the surface correction is given as
\begin{equation}
\nu_{corr}(k,l)=\nu_{best}(k,l)+(\frac{a}{\theta})[\frac {\nu_{obs}(k,l)}{\nu_{0}}]^{b}.
\end{equation}
Frequencies of the mixed mode are less affected by the surface effects (Kjeldsen et al. 2008), and the correction term is computed by the following equation (Brand{\~a}o et al. 2011, Tian et al. 2015)
\begin{equation}
\nu_{corr}(k,l)=\nu_{best}(k,l)+(\frac{a}{\theta})(\frac{1}{Q_{kl}})[\frac {\nu_{obs}(k,l)}{\nu_{0}}]^{b}.
\end{equation}
In Equation (9) and Equation (10), $\nu_{obs}(k,l)$ and $\nu_{best}(k,l)$ correspond to observed frequencies and our best-model frequencies. The value of the power index $b$ = 4.90 was given by Kjeldsen et al. (2008) in the case of the sun, and we also adopt this value in our study. The term $\nu_{0}$ is a reference frequency, and Kjeldsen et al. (2008) chose the solar maximum power frequency $\nu_{max}$ as the value of $\nu_{0}$. Following the method of Kjeldsen et al. (2008), we take the value of the maximum power frequency $\nu_{max}$ listed in Table 2 as the reference frequency: $\nu_{0}$ = 131.7 $\mu$Hz. As the constants of the stellar model, $\theta$ and $a$ can be calculated following the procedure of Kjeldsen et al. (2008)
\begin{equation}
\theta=(b-1)[b\frac{\langle\nu_{best}(k)\rangle}{\langle\nu_{obs}(k)\rangle}-\frac{\langle\Delta\nu_{best}(k)\rangle}{\langle\Delta\nu_{obs}(k)\rangle}]^{-1}
\end{equation}
and
\begin{equation}
a=\frac{\langle\nu_{obs}(k)\rangle-\theta\langle\nu_{best}(k)\rangle}{K^{-1}\sum_{i=1}^{K}[\nu_{obs}(k_{i})/\nu_{0}]^{b}}.
\end{equation}
Here, $\langle\nu_{obs}(k)\rangle$ and $\langle\nu_{best}(k)\rangle$ are the means of the observed frequencies and best-model frequencies respectively,
$\langle\Delta\nu_{obs}(k)\rangle$ and $\langle\Delta\nu_{best}(k)\rangle$ are the slope of a linear least-squares fit to the given frequencies and are
defined by
\begin{equation}
\langle\Delta\nu_{obs}(k)\rangle=\frac{\sum_{i=1}^{K}[\nu_{obs}(k_{i})-\langle\nu_{obs}(k)\rangle](k_{i}-\langle k\rangle)}{\sum_{i=1}^{K}(k_{i}-\langle k\rangle)^{2}}
\end{equation}
and
\begin{equation}
\langle\Delta\nu_{best}(k)\rangle=\frac{\sum_{i=1}^{K}[\nu_{best}(k_{i})-\langle\nu_{best}(k)\rangle](k_{i}-\langle k\rangle)}{\sum_{i=1}^{K}(k_{i}-\langle k\rangle)^{2}}.
\end{equation}
We have considered Model N as the best-fitting model in the last section. As a result, we obtain the values of $\theta$ and $a$ are: $\theta = 1.004$, $a = -0.354$. The correction term $Q_{kl}$ is the ratio between the inertia of nonradial mode $I_{kl}$ and the inertia of radial mode $I_{0}$ of the same
frequency (Aerts et al. 2010). $I_{kl}$ is defined in Equation (3). The ratio $Q_{kl}$ is presented as
\begin{equation}
Q_{kl}=\frac{I_{kl}}{I_{0}}\simeq1+\frac{l(l+1)\int_{0}^{R}|\xi_{h}(r)|^{2}\rho_{0}r^{2}dr}{\int_{0}^{R}|\xi_{r}(r)|^{2}\rho_{0}r^{2}dr}.
\end{equation}
Because the mixed modes are less affected by the surface effect compared with p modes as discussed by Kjeldsen et al. (2008) and the observed mixed modes
are highly consistent with the calculated mixed modes. Therefore, we only use surface effect corrections for the $l = 0$ and $l = 2$ modes.
The \'{e}chelle diagrams of the best-fitting model after applying the surface effect correction of $l = 0$ and $l = 2 $ frequencies are displayed in
Figure 11. We can see that after applying the corrections of surface effect, the agreement between observed and calculated frequencies is better than before. This result confirm that the offsets between observed and calculated frequencies of $l = 0$ and $l = 2$ are actually caused by the surface effect rather than the model calculations.

\section{Comparison with previous work}
Montalban et al. (2013) found that for low-mass red giant branch models the period spacing $\Delta P$ decreases as the helium core mass $M_{\rm He}$ increases. From their work, the corresponding period spacing $\Delta P$ is about 60 s when the helium core mass reaches 0.2 $M_{\odot}$. In our work, the helium core mass of KIC 9145955 is determined to be $M_{\rm He}$ = 0.210 $\pm$ 0.002 $M_{\odot}$ and the value of observed period spacing is given as $\Delta P$ = 77.01 s. In the work of Montalban et al. (2013), they constructed a series of red giant models with masses 0.9, 1.0, 1.5, 1.6, and 1.7 $M_{\odot}$ and chemical composition $Z$ = 0.02 and $Y$ = 0.278. We obtain the fundamental parameters of KIC 9145955 as $M$ = 1.24 $M_{\odot}$, $Z$ = 0.009 and $Y$ = 0.259. The relation between the period spacing $\Delta P$ and the mass of the helium core $M_{\rm He}$ found by Montalban et al. (2013) gives a lower value of the period spacing than in our work and this difference can be explained by different chemical compositions of stellar models. In addition, Montalban et al. (2013) presented that period spacings of their red giant models range from 20 s to 100 s and the masses of the helium core do not exceed 0.03 $M_{\odot}$. Therefore, the helium core mass we obtain are consistent with the relation between the helium core mass and period spacing proposed by Montalban et al. (2013).

Corsaro et al. (2015b) presented the evidence of clear acoustic glitch and measured the acoustic depth of the second helium ionization zone in 18 low luminosity red giants, including KIC 9145955, by using $l = 0$ and $l = 2$ modes. In our work, we use $l = 0$, $l = 2$ and the most p-dominated modes of $l = 1$ to calculate the first frequency difference $\Delta\nu_{n,l}$ and second frequency difference $\Delta_{2}\nu_{n,l}$ (e.g., Gough 1990) which are defined as
\begin{equation}
\Delta\nu_{n,l}\equiv\nu_{n+1,l}-\nu_{n,l}
\end{equation}
and
\begin{equation}
\Delta_{2}\nu_{n,l}\equiv\nu_{n+1,l}-2\nu_{n,l}+\nu_{n-1,l}.
\end{equation}
The results of $\Delta\nu_{n,l}$ and $\Delta_{2}\nu_{n,l}$ are shown in Figure 12. The results of Corsaro et al. show that the acoustic glitches of $l = 0$ and $l = 2$ are almost the same. In Figure 12, we can see clearly the acoustic glitches of $l = 0$, $l = 2$ and the most p-dominated $l = 1$ modes. Our results show that acoustic glitches with $l = 2$ and the most p-dominated $l = 1$ modes are very similar. In the top and bottom panel of Figure 12, the first and second frequency difference with the corresponding polynomial fits of the most p-dominated $l = 1$ modes are similar to $l = 2$ modes. But $l = 0$ modes are bumped slightly in the top and bottom panel of Figure 12.

P{\'e}rez Hern{\'a}ndez et al. (2016) obtained the fundamental parameters of KIC 9145955 by using the $l = 0$, $l = 2$ and $l = 3$ modes, the period spacings of mixed modes and spectroscopic data. Therefore the $l = 1$ modes were not fitted in their work. Their stellar parameters of KIC 9145955 are: $M$ = 1.196 $M_{\odot}$, $Z$ = 0.009, $Y$ = 0.294, $\alpha$ = 1.941, $f_{\rm ov}$ = 0.021, $R$ = 5.543 $R_{\odot}$, and $L$ = 18.496 $L_{\odot}$. Our fundamental parameters are consistent with the results of P{\'e}rez Hern{\'a}ndez et al. (2016) except for the initial helium abundance and overshooting parameter. We do not consider the convective overshooting. P{\'e}rez Hern{\'a}ndez et al. (2016) also mentioned that their initial helium abundance was not be determined reliably. Our initial helium abundance $Y$ = 0.259 is significantly lower than the value of P{\'e}rez Hern{\'a}ndez et al. (2016), but the value of $(Z/X)_{star}$ obtained by us is consistent with the spectroscopic data. Furthermore, we obtain the precise size of the helium core of KIC 9145955 through high-precision mixed modes: $M_{\rm He}$ = 0.210 $M_{\odot}$ and $R_{\rm He}$ = 0.0307 $R_{\odot}$. In summary, our stellar parameters are in agreement with the results of Perez Hernandez et al. (2016) and we obtain the parameters of helium core by fitting mixed modes individually.
\section{Conclusion}
In this work, we have analyzed high-quality photometric data of KIC 9145955 provided by the NASA \emph{Kepler} mission and obtained 61
individual oscillation frequencies. Then we carry out asteroseismic modeling for KIC 9145955. Some meaningful results are summarized as follows:

 1. The observed frequency F37 = 96.931 $\mu$Hz is identified as a mode of the most p-dominated from the period \'{e}chelle diagram of $l = 1$ modes. But our model calculations show that F39 = 96.397 $\mu$Hz has a higher p-mode oscillation energy ratio than F37 = 96.931 $\mu$Hz. In addition, if we certify F37 = 96.931 $\mu$Hz as a mode of the most p-dominated, we will not able to choose the best-fitting model which has the minimal value of $\chi_{k,l=1}^{2}$ and also shows a good agreement between observed and calculated $l = 1$ modes in frequency \'{e}chelle diagram. Our results confirm that it is more appropriate to identify F39 = 96.397 $\mu$Hz as a mixed mode of the most p-dominated.

 2. Through 38 high-precision g-dominated mixed modes, the size of the helium core of KIC 9145955 is determined to be
 $M_{\rm He}$ = 0.210 $\pm$ 0.002 $M_{\odot}$ and $R_{\rm He}$ = 0.0307 $\pm$ 0.0002 $R_{\odot}$. If we only use the
 asteroseismic results, the global parameters like stellar radius $R$, surface gravity log$g$ and luminosity $L$ will not be measured precisely except acoustic radius $\tau_{0}$ which reflects the characteristic of stellar envelope. The acoustic radius of KIC 9145955 is determined to be
 $\tau_{0}$ = 0.494 $\pm$ 0.001 days. We can obtain the best-fitting model by using asteroseismic results and spectroscopic limitation. The fundamental parameters of KIC 9145955 are determined to be $M$ = 1.24 $M_{\odot}$, $Z$ = 0.009, $\alpha$ = 2.0, $T_{eff}$ = 5069 K, $\log g$ = 3.029, $R$ = 5.636 $R_{\odot}$, and $L$ = 18.759 $L_{\odot}$.

 3. The offsets between observed and calculated frequencies, especially for p modes, are caused by the surface effect.
 After application of the empirical correction method proposed by Kjeldsen et al. (2008), we improve the agreement between observed p-mode frequencies and corresponding calculated ones.

4. In our model fittings, the mixed modes of $l = 1$ are fitted individually. Therefore, we obtain not only the fundamental parameters of KIC 9145955 but also the information about the helium core. The helium core mass we derived are in agreement with the conclusion of Montalban et al. (2013). We find the acoustic glitches of $l = 0$, $l = 2$ and the most p-dominated $l = 1$ modes.

\acknowledgments
This work is supported by the NSFC of China (Grant No.11333006,11521303 and 11503076), by the foundation of Chinese Academy of Sciences (Grant No. XDB09010202) and by Yunnan Applied Basic Research Projects (Grant No.2017B008).
The authors acknowledge the NASA and \emph{Kepler} team for allowing them to work with and analyze the \emph{Kepler} data. The \emph{Kepler} Mission is funded by NASA¡¯s Science Mission Directorate. The authors gratefully acknowledge the computing time granted by the Yunnan Observatories, and provided on the facilities at the Yunnan Observatories Supercomputing Platform. The authors sincerely thank the instructive comments and helpful suggestions from an anonymous referee. The authors also acknowledge the suggestions from X.-H. Chen and J.-J. Guo.

\begin{figure}
\centering
\includegraphics[angle=-90,scale=0.4]{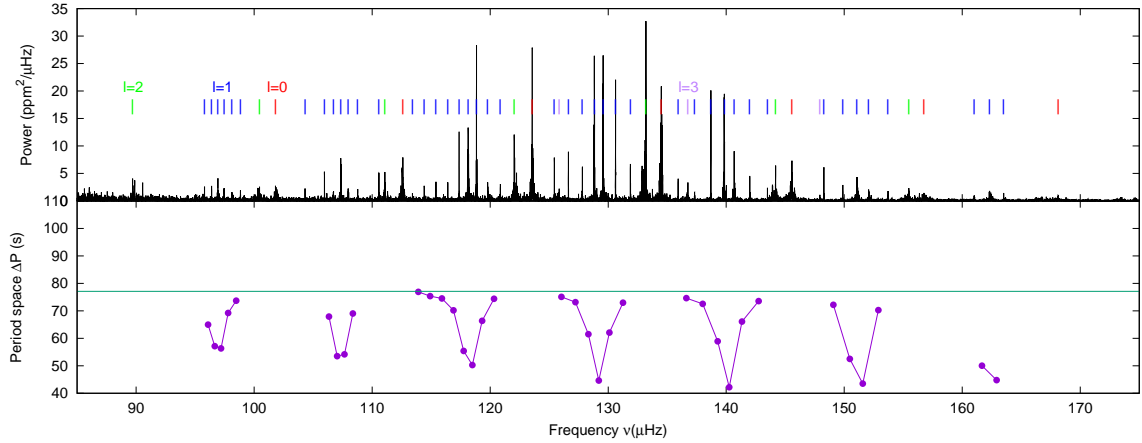}
\caption{Top: Power density spectrum of KIC 9145955. Each colored line represents an individual frequency. Bottom: Each filled cycle represents a period spacing between two adjacent mixed modes. The horizontal line indicates the period spacing of the g-dominated mixed modes.}
\label{Figure.1}
\end{figure}

\begin{figure}
\centering
\includegraphics[angle=0,scale=1.0]{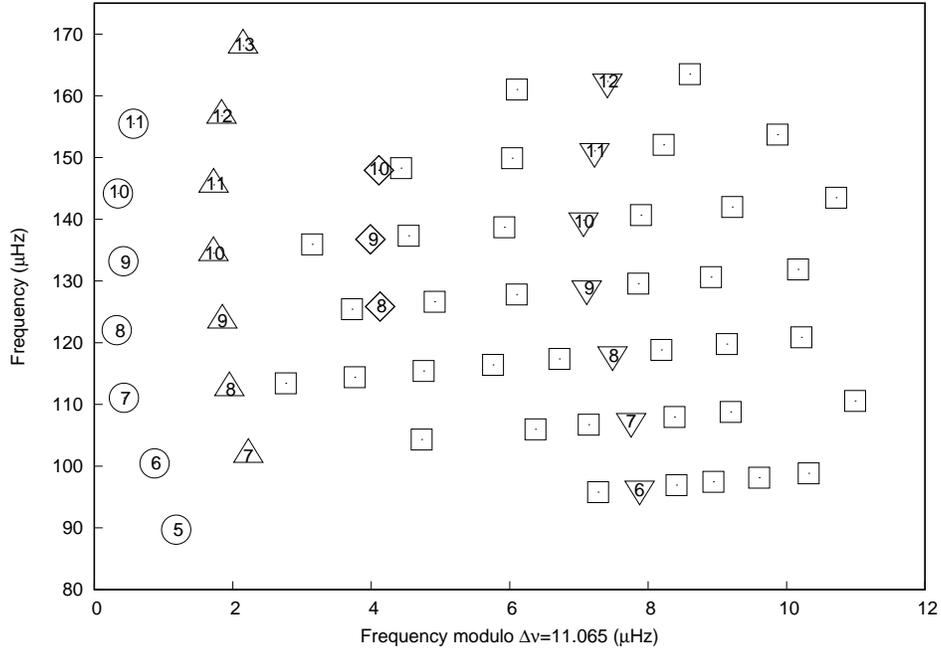}
\caption{Frequency \'{e}chelle diagram for KIC 9145955. The triangles, circles, and diamonds denote the $l = 0 $, $l = 2$, and $l = 3$ modes.
The squares denote $l = 1$ mixed modes, but with an inverted triangle in each order showing the $l = 1$ mode that is the most p-dominated. The figures in open symbols indicate the radial order (n) of the modes which are defined by model calculations.}
\label{Figure.2}
\end{figure}

\begin{figure}
\centering
\includegraphics[angle=0,scale=1.0]{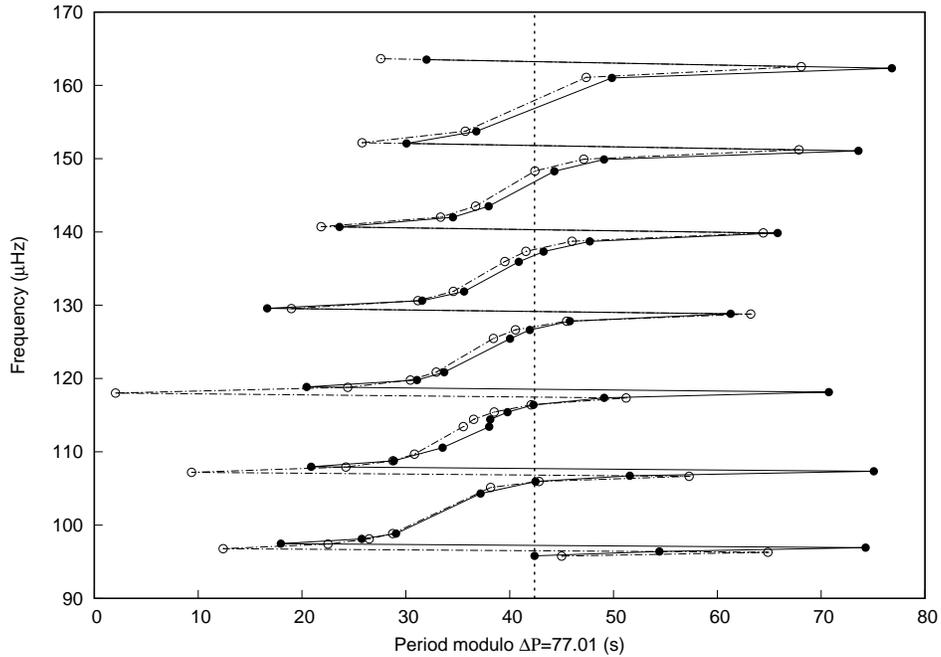}
\caption{Period \'{e}chelle diagram of $l = 1$ modes for KIC 9145955. The filled circles denote observed $l = 1$ modes and the open circles denote calculated $l = 1$ modes. The vertical dashed line denotes the vertical alignment of the high order g-dominated mixed modes. The mode identification see section 4.1.}
\label{Figure.3}
\end{figure}

\begin{figure}
\includegraphics[angle=0,scale=1.5]{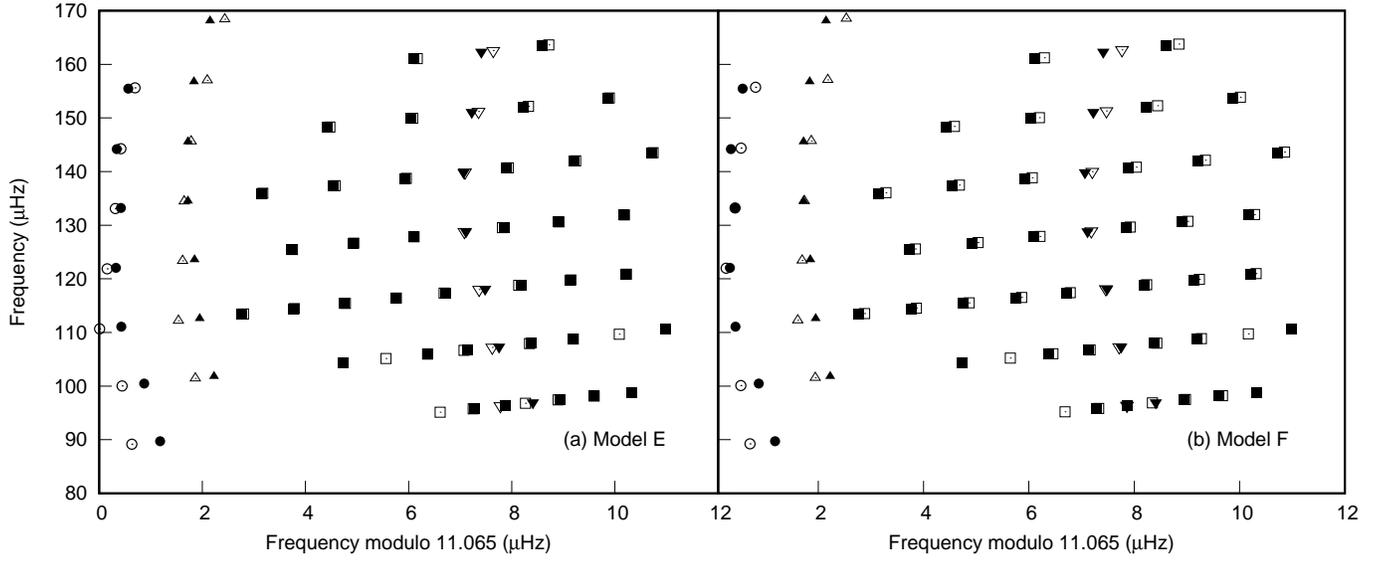}
\caption{\'{E}chelle diagrams of observed frequencies and corresponding calculated frequencies. Panel (a) and panel (b) show the \'{e}chelle diagram of
Model E and Model F, respectively. The filled and open symbols denote observed and calculated frequencies. The triangles and circles denote $l = 0$ modes and $l = 2$ modes, respectively. The squares denote $l = 1$ mixed modes, but with an inverted triangle in each order showing the $l = 1$ mode that is the most p-dominated.
The parameters of models are listed in Table 3.}
\label{Figure.4}
\end{figure}

\begin{figure}
\includegraphics[angle=0,scale=1.0]{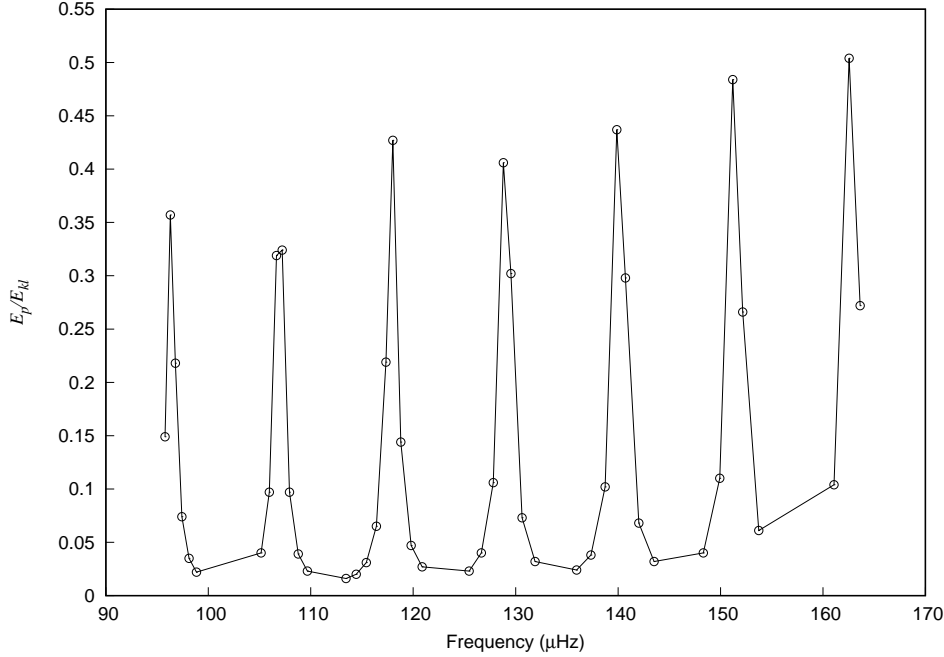}
\caption{The energy ratio of p-mode for each calculated $l = 1$ mode. Each cycle represents a calculated $l = 1$ mode.}
\label{Figure.5}
\end{figure}

\begin{figure}
\includegraphics[angle=0,scale=1.5]{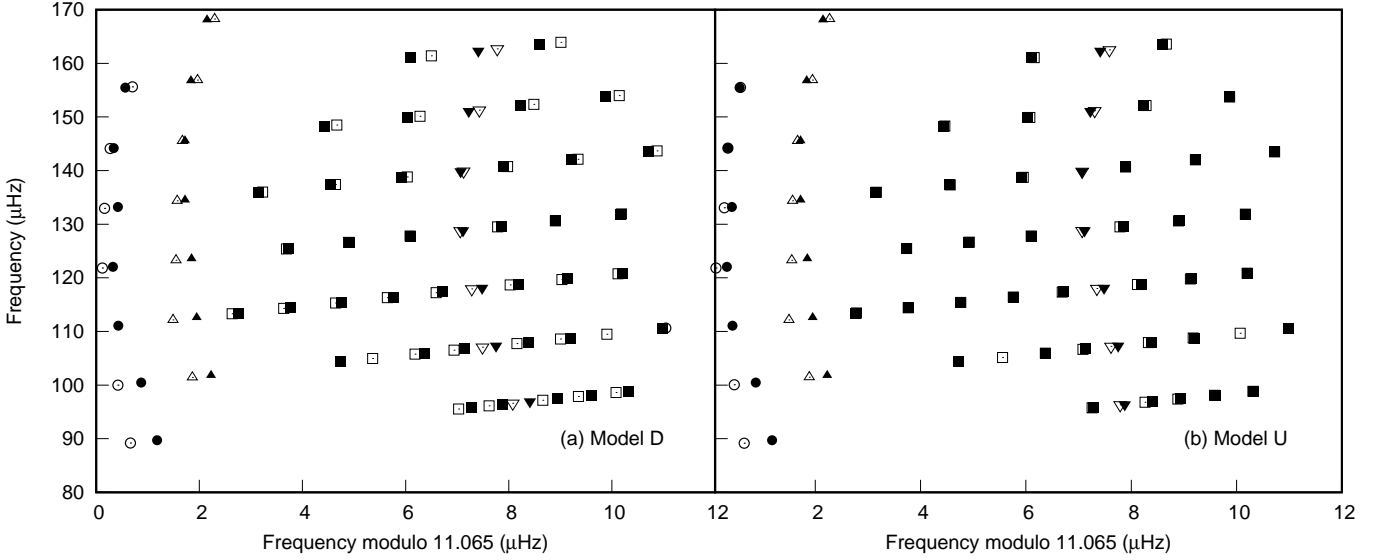}
\caption{\'{E}chelle diagrams of observed frequencies and corresponding calculated frequencies. The filled and open symbols denote observed and calculated
frequencies. The triangles and circles denote $l = 0$ modes and $l = 2$ modes. The squares denote $l = 1$ mixed modes, but with an inverted triangle in each order showing the $l = 1$ mode that is the most p-dominated. Panel (a) shows the \'{e}chelle diagram of Model D which considers F37 = 96.931 $\mu$Hz as the most p-dominated mode. Panel (b) shows the \'{e}chelle diagram of Model U which considers F39 = 96.397 $\mu$Hz as the most p-dominated mode. The parameters of Model D and Model U are listed in Table 3 and Table 4, respectively.}
\label{Figure.6}
\end{figure}

\begin{figure}
\centering
\includegraphics[angle=0,scale=1.0]{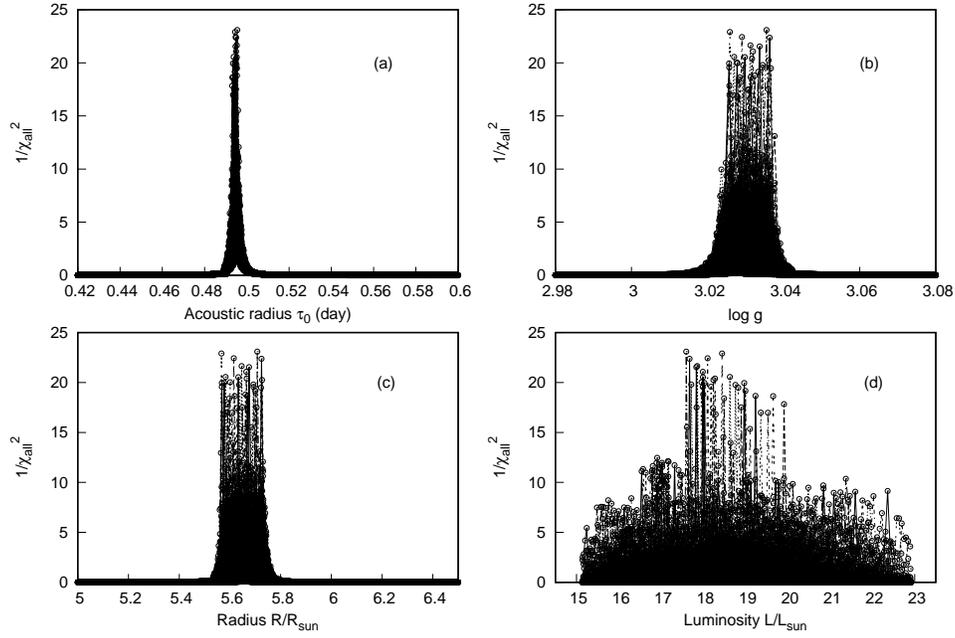}
\caption{$\chi^{2}_{\rm all}$ as a function of acoustic radius $\tau_{0}$ (panel(a)), surface gravity $\log g$ (panel (b)), stellar radius $R$ (panel (c)), and stellar luminosity $L$ (panel (d)) for all 550 evolutional tracks.}
\label{Figure.7}
\end{figure}

\begin{figure}
\includegraphics[angle=0,scale=1.5]{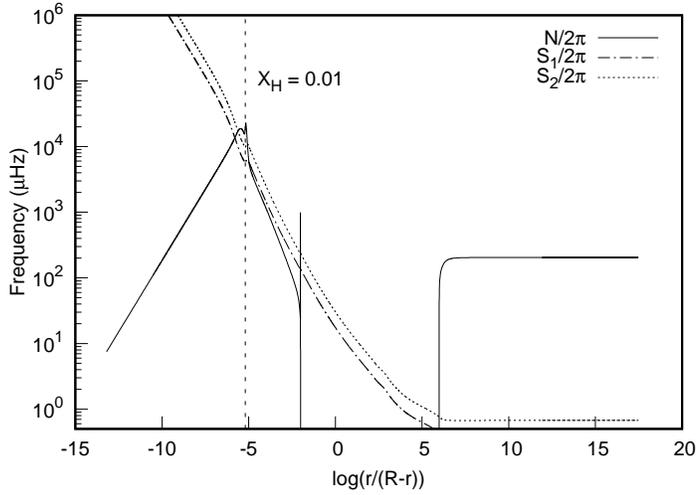}
\caption{Propagation diagram of Model N. The continues line indicates Buoyancy frequency $N$ and the dashed line indicates characteristic
acoustic frequency $S_{1,2}$. The vertical dashed line denotes the helium core boundary where the hydrogen mass fraction is about 0.01. $R$ is the stellar radius. }
\label{Figure.8}
\end{figure}

\begin{figure}
\includegraphics[angle=0,scale=1.0]{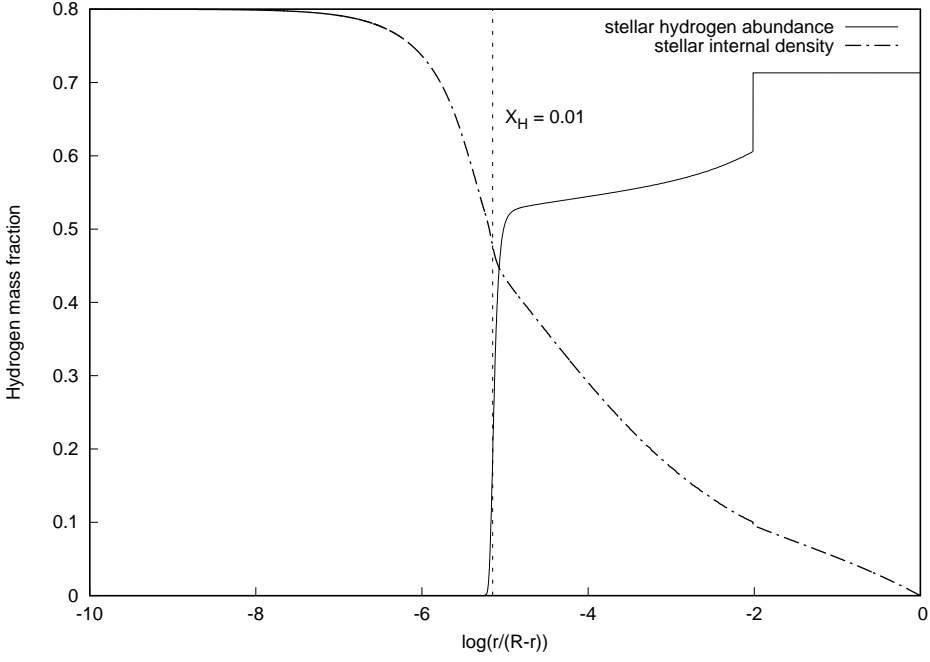}
\caption{Hydrogen abundance and internal density of Model N. The vertical dashed line denotes the boundary of helium core. The density line with a logarithmic ordinate and the hydrogen abundance line have the same abscissa in the graph. The slope of density line changes at the boundary of helium core.}
\label{Figure.9}
\end{figure}

\begin{figure}
\includegraphics[angle=0,scale=1.5]{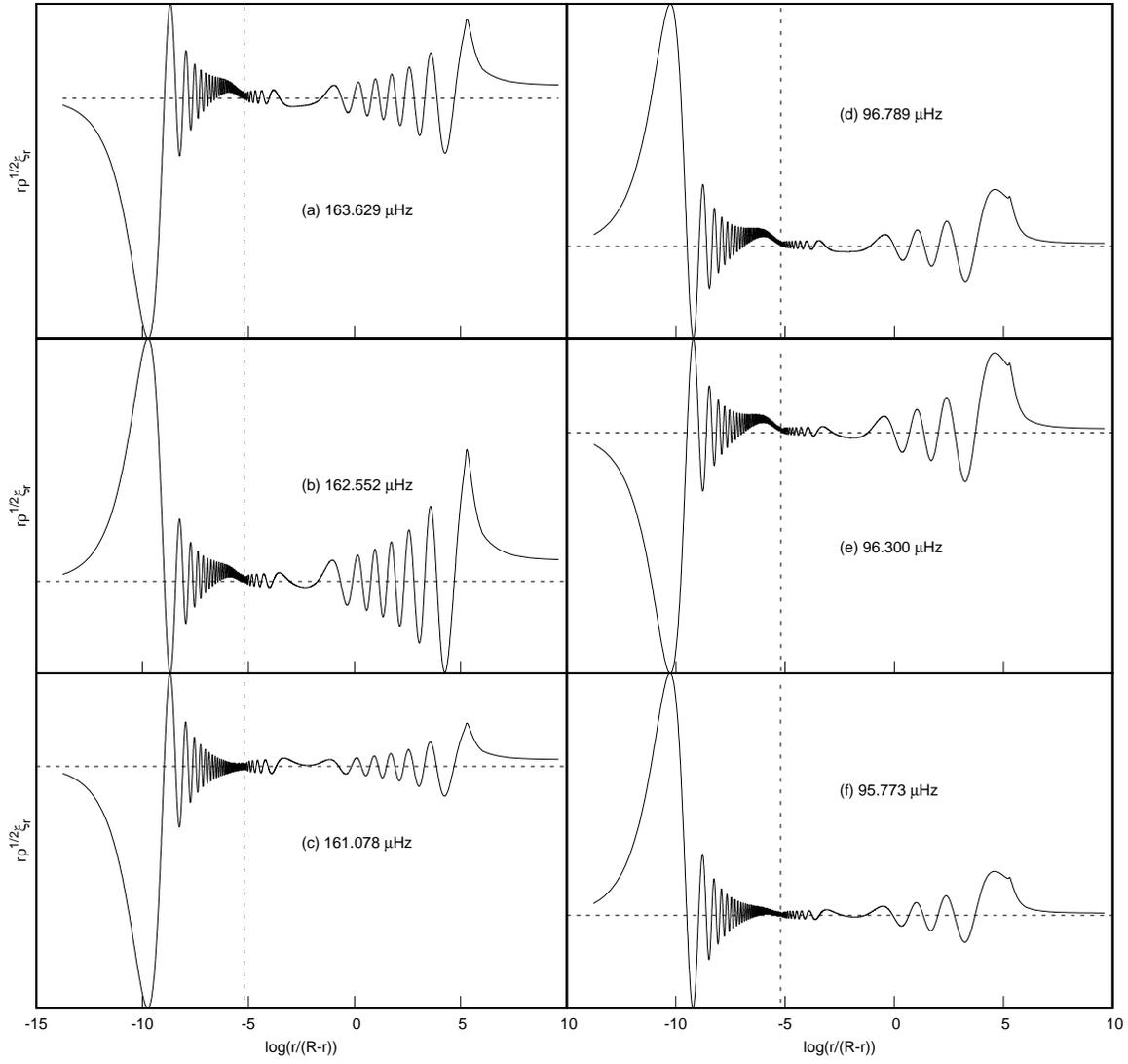}
\caption{Scaled displacement eigenfunctions of six calculated frequencies for Model N. The vertical dashed line is the boundary of helium core. The information of these frequencies are listed in Table 5.}
\label{Figure.10}
\end{figure}

\begin{figure}
\includegraphics[angle=0,scale=1.5]{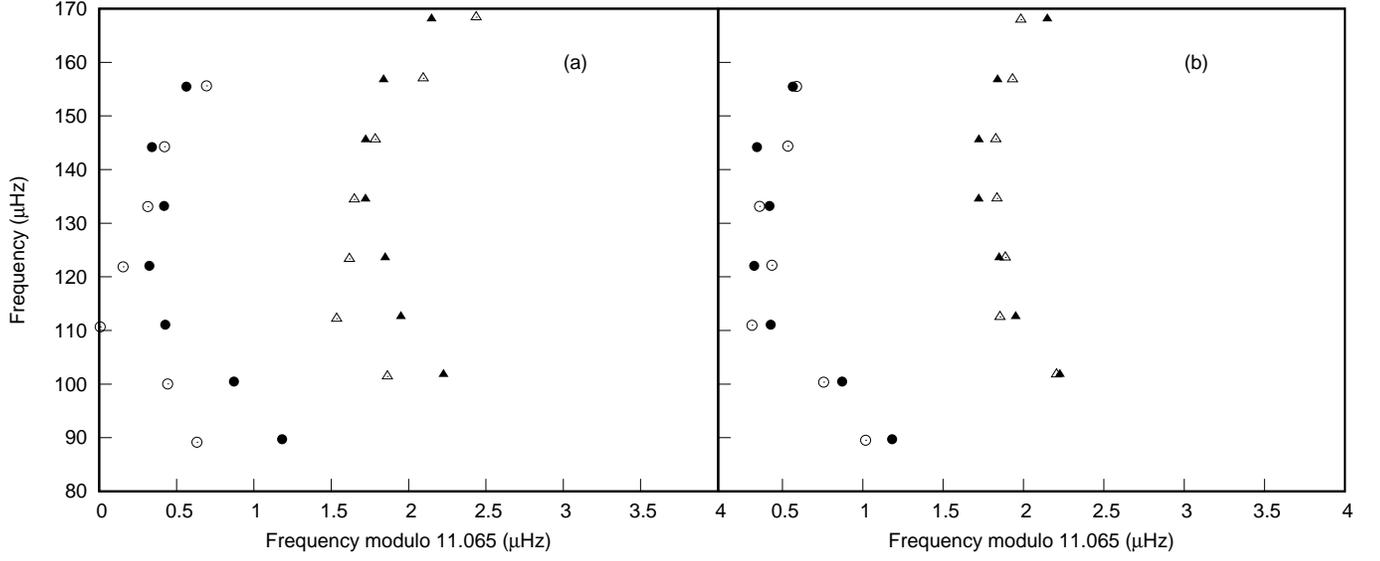}
\caption{\'{E}chelle diagrams of observed $l$ = 0 and 2 frequencies and corresponding calculated frequencies for the best-fitting model. Panel (a) and panel (b) show before and after application of the surface effect corrections to the calculated frequencies. The filled and open symbols denote observed and calculated frequencies, respectively. The triangles, circles denote $l = 0$ modes, $l = 2$ modes.}
\label{Figure.11}
\end{figure}

\begin{figure}
\includegraphics[angle=0,scale=1.4]{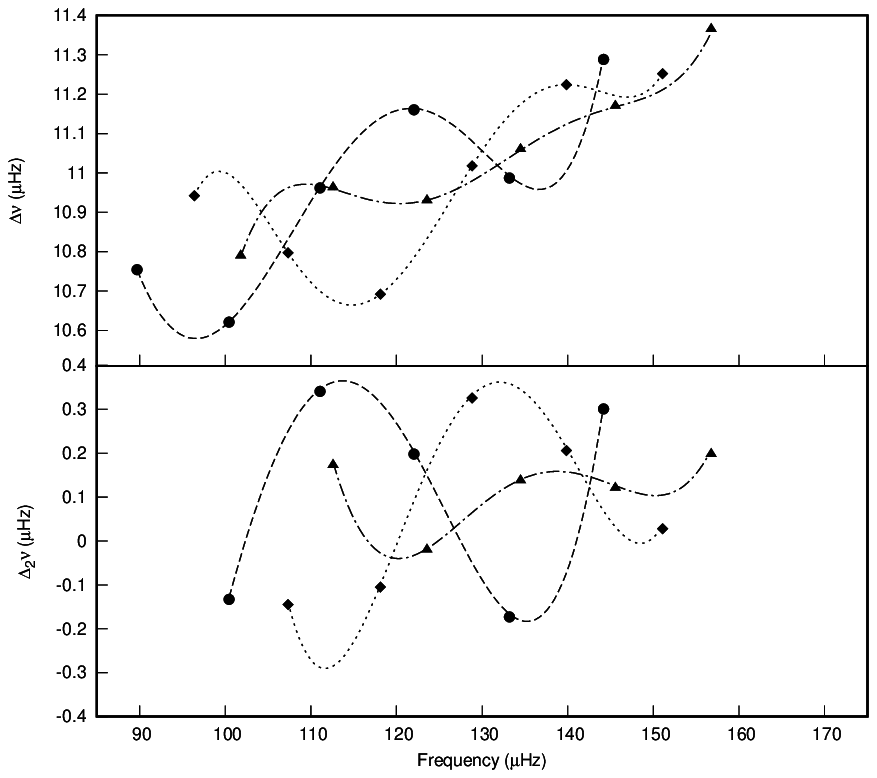}
\caption{The first and second difference of modes. Top: the first difference, $\Delta\nu_{n,l}$, Eq.(16). The triangles and circles represent values  computed from $l = 0$  and $l = 2$ modes. The corresponding polynomial fit is indicated by dash-dot and dashed lines, respectively. The diamonds represent values computed from $l = 1$ modes that are the most p-dominated. The corresponding polynomial fit is indicated by dotted line. Bottom: the second difference, $\Delta_{2}\nu_{n,l}$, Eq.(17). The triangles and circles represent values computed from $l = 0$  and $l = 2$ modes. The corresponding polynomial fit is indicated by dash-dot and dashed lines, respectively. The diamonds represent values computed from $l = 1$ modes that are the most p-dominated. The corresponding polynomial fit is indicated by dotted line.}
\label{Figure.12}
\end{figure}

\begin{table*}
\caption{\label{t1}Oscillation frequencies of KIC 9145955 extracted by Period04. The ID is the serial number of observed frequencies in the Figure 1, which represents the peak number arranges from left to right in the power density spectrum. The serial number of observed frequencies in Corsaro et al. (2015a) represents the peak number in increasing frequency order.}
\centering
\begin{tabular}{cccccccccc}
\hline\hline
$l$   &ID    &Frequency   &Uncertainty   &Amplitude          &$l$  &ID   &Frequency   &Uncertainty    &Amplitude      \\
      &      &($\mu$Hz)   &($\mu$Hz)     &(ppm)              &     &     &($\mu$Hz)   &($\mu$Hz)      &(ppm)           \\
\hline
0     &F35   &101.811     &0.032         &5.593              &1    &F18  &127.815     &0.194          &8.503                  \\
0     &F26   &112.601     &0.012         &9.515              &1    &F4   &128.828     &0.0102         &17.360                 \\
0     &F3    &123.564     &0.012         &18.067             &1    &F5   &129.573     &0.0102         &17.029                \\
0     &F7    &134.501     &0.090         &15.747             &1    &F6   &130.624     &0.231          &15.889               \\
0     &F47   &145.568     &0.012         &9.402              &1    &F45  &131.881     &0.012          &8.837                \\
0     &F54   &156.749     &0.012         &4.086              &1    &F17  &135.928     &0.012          &6.733               \\
0     &F58   &168.124     &0.231         &3.405              &1    &F15  &137.321     &0.088          &4.052               \\
1     &F40   &95.797      &0.012         &5.455              &1    &F8   &138.703     &0.231          &15.298               \\                                                                              1     &F39   &96.397      &0.045         &5.532              &1    &F9   &139.846     &0.231          &15.292               \\
1     &F37   &96.931      &0.029         &6.727              &1    &F13  &140.676     &0.154          &10.383               \\
1     &F38   &97.463      &0.231         &5.023              &1    &F14  &141.996     &0.012          &7.443                \\
1     &F41   &98.125      &0.012         &4.359              &1    &F59  &143.495     &0.231          &5.329                \\
1     &F42   &98.840      &0.012         &4.563              &1    &F48  &148.276     &0.026          &8.482                \\
1     &F34   &104.314     &0.036         &5.170              &1    &F49  &149.881     &0.085          &5.849                \\
1     &F33   &105.958     &0.012         &7.776              &1    &F50  &151.070     &0.051          &7.096                 \\
1     &F32   &106.726     &0.113         &4.672              &1    &F51  &152.070     &0.044          &4.670                 \\
1     &F31   &107.339     &0.012         &9.297              &1    &F52  &153.713     &0.012          &4.568                  \\
1     &F30   &107.967     &0.012         &5.135              &1    &F55  &161.014     &0.031          &3.407                  \\
1     &F29   &108.778     &0.093         &4.854              &1    &F56  &162.322     &0.231          &4.730                 \\
1     &F28   &110.574     &0.012         &7.636              &1    &F57  &163.511     &0.012          &4.055                 \\
1     &F62   &113.415     &0.012         &4.003              &2    &F44  &89.702      &0.012          &6.825                  \\
1     &F25   &114.413     &0.143         &5.702              &2    &F36  &100.456     &0.012          &5.304                  \\
1     &F24   &115.408     &0.168         &6.235              &2    &F27  &111.077     &0.012          &7.559                   \\
1     &F23   &116.409     &0.012         &6.460              &2    &F12  &122.039     &0.012          &11.858                  \\
1     &F10   &117.368     &0.056         &12.140             &2    &F1   &133.199     &0.012          &19.703                  \\
1     &F11   &118.136     &0.012         &12.404             &2    &F46  &144.186     &0.231          &8.899                   \\
1     &F2    &118.842     &0.036         &17.959             &2    &F53  &155.474     &0.110          &5.268                   \\
1     &F22   &119.787     &0.231         &6.246              &3    &F60  &125.850     &0.195          &5.166                   \\
1     &F21   &120.864     &0.231         &5.835              &3    &F16  &136.755     &0.012          &6.325                   \\
1     &F20   &125.438     &0.161         &9.545              &3    &F61  &147.937     &0.036          &3.458                   \\
1     &F19   &126.631     &0.012         &10.272       \\
\hline
\end{tabular}
\end{table*}

\begin{table*}
\caption{\label{t2}Spectroscopic parameters of the red giant KIC 9145955.}
\centering
\begin{tabular}{ccccccccc}
\hline\hline
$T_{\rm eff}$ [K]  &log$g$ [dex]  &$\rm [Fe/H]$ &$\Delta\nu$ [$\mu$Hz] &$\nu_{\rm max}$ [$\mu$Hz]   &$\Delta \rm P$ [s] &Reference\\
4943                 &2.85            &-0.34            &11.00             &130.0                   &77.01            &Takeda et al.(2016) \\
4925 $\pm$ 91        &3.04 $\pm$ 0.11 &-0.32 $\pm$ 0.03 &11.00 $\pm$ 0.06  &131.7 $\pm$ 0.2         &                &P{\'e}rez Hern{\'a}ndez et al.(2016) \\
&                    &                                 &11.065            &                        &77.01            &In this paper \\
\hline
\end{tabular}
\end{table*}

\begin{table*}
\caption{\label{t3}The experimental models with the mixing-length $\alpha$ = 2.0. In this table, we identify observed frequency F37 = 96.931 $\mu$Hz as the most p-dominated mode. The last two columns are the mass and radius of the helium core, respectively.}
\centering
\begin{tabular}{ccccccccccccccccc}
\hline\hline
Model   &$\alpha$  &initial $M$   &initial $Z$  &$T_{\rm eff}$ &log$g$ &$R$           &$L$            &$\chi^{2}_{\rm all}$ &$\chi^{2}_{\rm l=1}$ &$M_{\rm
He}$   &$R_{\rm He}$  \\
&&$[M_{\odot}]$ &&[K]&[dex]&$[R_{\odot}]$ &$[L_{\odot}]$ & & &[$M_{\odot}$] &[$R_{\odot}$] \\
\hline
A       &2.00      &1.20  &0.012   &4986          &3.025  &5.574          &17.172         &0.0984     &0.0874    &0.208    &0.0308\\
B       &2.00      &1.21  &0.013   &4971          &3.027  &5.584          &17.021         &0.0770     &0.0776    &0.208    &0.0308\\
C       &2.00      &1.22  &0.014   &4955          &3.028  &5.600          &16.908         &0.0775     &0.0782    &0.207    &0.0309\\
D       &2.00      &1.23  &0.015   &4940          &3.029  &5.618          &16.803         &0.0786     &0.0742    &0.207    &0.0310\\
E       &2.00      &1.24  &0.009   &5069          &3.029  &5.636          &18.759         &0.0966     &0.0992    &0.210    &0.0307\\
F       &2.00      &1.25  &0.009   &5073          &3.031  &5.649          &18.894         &0.0933     &0.0988    &0.210    &0.0307\\
G       &2.00      &1.26  &0.012   &5004          &3.032  &5.664          &17.989         &0.0904     &0.0960    &0.210    &0.0308\\
H       &2.00      &1.27  &0.013   &4988          &3.034  &5.676          &17.834         &0.0853     &0.0957    &0.209    &0.0308\\
\hline
\end{tabular}
\end{table*}

\begin{table*}
\caption{\label{t4}The candidate models for KIC 9145955. In this table, we identify the observed frequency F39 = 96.397 $\mu$Hz as the most p-dominated mode. The last column is the acoustic radius computed by the asymptotic theory of stellar oscillations.}
\centering
\begin{tabular}{ccccccccccccccccc}
\hline\hline
Model   &$\alpha$  &initial $M$   &initial $Z$   &$T_{\rm eff}$ &log$g$ &$R$  &$L$  &$\chi^{2}_{\rm all}$ &$\chi^{2}_{l=1}$ &$M_{\rm He}$   &$R_{\rm He}$
&$Z/X$  &$\tau_{0}$\\
&&$[M_{\odot}]$ &&[K]&[dex]&$[R_{\odot}]$ &$[L_{\odot}]$ & & &[$M_{\odot}$] &[$R_{\odot}$]& &[day] \\
\hline
I       &1.90      &1.26  &0.007   &5086          &3.031  &5.699          &19.234         &0.0536     &0.0410    &0.211    &0.0306   &0.0109 &0.494\\
J       &1.90      &1.28  &0.008   &5056          &3.033  &5.700          &18.996         &0.0522     &0.0398    &0.211    &0.0306   &0.0123 &0.494\\
K       &1.90      &1.30  &0.012   &4955          &3.036  &5.724          &17.673         &0.0447     &0.0403    &0.210    &0.0307   &0.0182 &0.495\\
L       &2.00      &1.21  &0.006   &5166          &3.026  &5.591          &19.909         &0.0561     &0.0396    &0.212    &0.0306   &0.0095 &0.493\\
M       &2.00      &1.23  &0.007   &5136          &3.028  &5.619          &19.653         &0.0537     &0.0406    &0.211    &0.0307   &0.0109 &0.493\\
N       &2.00      &1.24  &0.009   &5069          &3.029  &5.636          &18.759         &0.0506     &0.0385    &0.210    &0.0307   &0.0139 &0.494\\
O       &2.00      &1.26  &0.011   &5023          &3.032  &5.666          &18.277         &0.0490     &0.0384    &0.210    &0.0307   &0.0168 &0.495\\
P       &2.00      &1.26  &0.012   &5004          &3.032  &5.664          &17.989         &0.0475     &0.0395    &0.210    &0.0308   &0.0183 &0.495\\
Q       &2.00      &1.29  &0.015   &4957          &0.035  &5.708          &17.596         &0.0433     &0.0410    &0.209    &0.0309   &0.0227 &0.496\\
R       &2.10      &1.20  &0.009   &5114          &3.026  &5.568          &18.966         &0.0501     &0.0407    &0.210    &0.0307   &0.0139 &0.493\\
S       &2.10      &1.21  &0.010   &5085          &3.027  &5.583          &18.628         &0.0486     &0.0410    &0.210    &0.0307   &0.0154 &0.494\\
T       &2.10      &1.23  &0.013   &5033          &3.029  &5.615          &18.099         &0.0446     &0.0386    &0.210    &0.0308   &0.0198 &0.495\\
U       &2.10      &1.24  &0.014   &5018          &3.030  &5.633          &17.990         &0.0487     &0.0381    &0.209    &0.0309   &0.0213 &0.495\\
V       &2.10      &1.25  &0.015   &5002          &3.031  &5.646          &17.854         &0.0462     &0.0412    &0.209    &0.0309   &0.0228 &0.495\\
W       &2.20      &1.20  &0.013   &5079          &3.026  &5.567          &18.442         &0.0436     &0.0392    &0.210    &0.0309   &0.0199 &0.494\\
\hline
\end{tabular}
\end{table*}

\begin{table*}
\caption{\label{t5}The information of the calculated frequencies for Model N in Figure 10.}
\centering
\begin{tabular}{ccccccccc}
\hline\hline
ID  &Observed Frequency   &Model Frequency  &$E_{p}/E_{kl}$ \\
    &[$\mu$Hz]            &[$\mu$Hz]        &                \\
\hline
F57  &163.511             &163.629          &0.272        \\
F56  &162.322             &162.552          &0.504        \\
F55  &161.014             &161.078          &0.104           \\
F37  &96.931              &96.789           &0.218           \\
F39  &96.397              &96.300           &0.357           \\
F40  &95.797              &95.773           &0.149            \\
\hline
\end{tabular}
\end{table*}

\end{document}